\documentclass[twocolumn,pra,showpacs,floatfix,amsmath,amsfonts]{revtex4-1}
\usepackage{multirow}
\usepackage[latin9]{inputenc}
\usepackage{color}
\usepackage{amstext}
\usepackage{graphicx}
\usepackage[table]{xcolor}

\makeatletter
\@ifundefined{textcolor}{}
{%
 \definecolor{BLACK}{gray}{0}
 \definecolor{WHITE}{gray}{1}
 \definecolor{RED}{rgb}{1,0,0}
 \definecolor{GREEN}{rgb}{0,1,0}
 \definecolor{BLUE}{rgb}{0,0,1}
 \definecolor{CYAN}{cmyk}{1,0,0,0}
 \definecolor{MAGENTA}{cmyk}{0,1,0,0}
 \definecolor{YELLOW}{cmyk}{0,0,1,0}
}

\usepackage{color}
\usepackage{graphics}
\usepackage{epsfig}
\newcommand{\be}{\begin{equation}}
\newcommand{\ee}{\end{equation}}
\newcommand{\bea}{\begin{eqnarray}}
\newcommand{\eea}{\end{eqnarray}}
\newcommand{\bes}{\begin{subequations}}
\newcommand{\ees}{\end{subequations}}

\newcommand{\tmu}{\tilde{\mu}}
\newcommand{\tom}{\tilde{\omega}}

\newcommand{\bPsi}{\mathbf{\Psi}}

\makeatother

\begin{document}

\title{Dynamic localization in optical and Zeeman lattices in the presence
of spin-orbit coupling}

\author{Yaroslav V. Kartashov$^{1,2}$, Vladimir V. Konotop$^{3}$, Dmitry
A. Zezyulin$^{3}$, and Lluis Torner$^{1,4}$ }

\affiliation{$^{1}$ ICFO-Institut de Ciencies Fotoniques, The Barcelona Institute
of Science and Technology, 08860 Castelldefels (Barcelona), Spain
\\
 $^{2}$Institute of Spectroscopy, Russian Academy of Sciences, Troitsk,
Moscow Region, 142190, Russia \\
 $^{3}$ Centro de F\'{i}sica Teórica e Computacional Faculdade de
Ciências and Departamento de F\'{i}sica, Faculdade de Ciências, Universidade
de Lisboa, Campo Grande, Ed. C8, Lisboa 1749-016, Portugal\\
 $^{4}$ Universitat Politecnica de Catalunya, 08034 Barcelona, Spain }

\date{\today}


\begin{abstract}
The dynamic localization of a two-level atom in a periodic potential under
the action of spin-orbit coupling and a weak harmonically-varying linear
force is studied. We consider optical and Zeeman potentials that are
either in-phase or out-of-phase in two spinor components, respectively.
The expectation value for the position of the atom after one oscillation
period of the linear force is recovered in authentic resonances or in pseudo-resonances.
The frequencies of the linear force corresponding to authentic resonances are determined
by the band structure of the periodic potential and are affected by
the spin-orbit coupling. The width/dispersion of the wavepacket in
authentic resonances is usually minimal. The frequencies corresponding to pseudo-resonances
do not depend on the type of potential and on the strength of the spin-orbit
coupling, while the evolution of excitations at the corresponding frequencies
is usually accompanied by significant dispersion. Pseudo-resonances are
determined by the initial phase of the linear force and by the quasi-momentum
of the wavepacket. Due to the spinor nature of the system, the motion
of the atom is accompanied by periodic, but not harmonic, spin oscillations.
 Under the action of spin-orbit coupling the oscillations
of the wavepacket can be nearly completely suppressed in optical lattices.
Dynamic localization in Zeeman lattices is characterized by  doubling
of the resonant oscillation periods due to band crossing at the boundary
of the Brillouin zone. We also show that higher harmonics in the Fourier
expansion of the energy band lead to effective dispersion, which can
be strong enough to prevent dynamic localization of the Bloch wavepacket.
\end{abstract}

\pacs{37.10.Jk, 32.60.+i, 03.75.-b}

\maketitle

\section{Introduction}

A linear force applied to a quantum particle in a periodic potential
does not lead to unidirectional motion, but results in particle
oscillations. This is the celebrated phenomenon of {\it Bloch oscillations\/}
known since the early days of quantum mechanics~\cite{Zener}. Bloch
oscillations have been experimentally observed in multiple solid-state settings~\cite{experim_solid}, in atomic systems~\cite{BEC,BEC_Bloch_theor},
and in optics~\cite{Peshel} (for reviews of the current state of the art, see~\cite{reviews}).
An oscillatory motion, or spatial localization, of a particle
in a periodic potential can also be caused by a linear force whose
amplitude oscillates with a properly selected frequency. This phenomenon,
known as {\it dynamic localization\/}, was discovered in Ref.~\cite{DunKenk}
within the framework of the tight-binding approximation widely used
in physics. It was later studied in systems of cold atoms held in optical
lattices (OLs)~\cite{atom-DL}, where it was used for the coherent
control of atoms and for realization of the superfluid -- Mott insulator
phase transition~\cite{coher_control}. Dynamic localization occurs also in optical systems \cite{optics-DL}, where a periodic
waveguide bending may emulate a time-periodic linear force acting on a
quantum particle. The effect can be used, e.g., to cause rectification of beam propagation trajectories \cite{rectification}.

Dynamic localization occurs also in nonlinear systems. In the theory of integrable systems, the dynamics of solitons in special
nonlinear lattices (including those governed by the Ablowitz-Ladik
model~\cite{AL}) subject to time-dependent linear forces was addressed in Ref.~\cite{BLR}
even before the phenomenon of dynamic localization caused by an oscillating
force. Actually, the linear
limit of the above mentioned model is the tight-binding
approximation used in Ref.~\cite{DunKenk} and in many subsequent studies
of dynamic localization. The oscillatory motion of solitons affected
by a time-periodic linear force as a manifestation of their dynamic localization
was discussed in~\cite{KCV}. The interplay between dynamic localization
and nonlinearity has been also studied in the meanfield theory of Bose-Einstein
condensates loaded in optical lattices, both theoretically~\cite{DL-theory}
and experimentally~\cite{DL-experiment}.

It should be properly appreciated that when compared with Bloch oscillations,
dynamic localization offers more opportunities for controlling the particle dynamics (or, in optics, the light propagation).
Indeed, it allows changing the direction of motion and average velocity of the particles
simply by varying the frequency of the linear force. In contrast,
a constant force leads only to oscillatory dynamics and the amplitude
of the force affects only the amplitude of the Bloch oscillations.

All studies mentioned above deal with dynamic localization of electrons,
atoms, or optical wavepackets described by a one-component wavefunction
or, alternatively, by a single quantum number (quasi-momentum). In
this paper we study dynamic localization of two-level atoms
characterized by a {\it spinor wavefunction\/}. The quantum numbers describing
such atom are the quasi-momentum and (pseudo-)spin.

The systems of
effective two-level atoms are attracting increasing attention due
to the proven experimental possibility of creation of synthetic electric~\cite{electric}
and magnetic~\cite{magnetic,magnetic_1} fields in them, which opens new avenues
for the investigation of the physics of gauge potentials~\cite{multi-level}.
One of the most significant advances in this direction is the creation
of spin-orbit (SO) coupled Bose-Einstein condensates (BECs)~\cite{Spielman}
{[}see \cite{SO-review} for a review{]}, where periodic potentials
can be induced not only in the form of OLs \cite{BEC}, equal for
both spinor components, but also in the form of Zeeman lattices (ZLs)
\cite{ZL_experiment}, which are out-of-phase for two spinor components.

Studies of Bloch oscillations of SO coupled atoms and BECs loaded
in OLs~\cite{Witthaut,Ke,Larson,KKZT} and in ZLs~\cite{Ke,KKZT},
have already been reported. Here we address the impact of a
time-periodic modulation of the amplitude of the linear force on the
dynamics of two-level SO-coupled atoms. The problem
and main analytical results are introduced in Sec.~\ref{sec:general}.
In Sections \ref{sec:OL} and \ref{sec:ZL} we describe the phenomenon
of dynamic localization for an atom loaded in an OL and in a ZL, respectively.
The results are summarized in the Conclusions section.

\section{Model}

\label{sec:general}

We consider a two-level atom characterized by a spinor $|\Psi\rangle=\boldsymbol{\Psi}=(\Psi_{1},\Psi_{2})^{T}$
(hereafter the superscript ``$T$'' stands for transposition).
Throughout the paper we deal with periodic potentials described by
the matrix
\begin{eqnarray}
h_{\epsilon}(x)=\epsilon V(x)I+(1-\epsilon)\Omega_{3}(x)\sigma_{3}.\label{HL}
\end{eqnarray}
Here $\sigma_{j}$ ($j=1,2,3$) are the standard Pauli matrices, $I$
is the $2\times2$ unit matrix, $\epsilon=1$ corresponds to an OL $h_{1}=V(x)I$,
while $\epsilon=0$ corresponds to a ZL $h_{0}(x)=\Omega_{3}(x)\sigma_{3}$.
The dimensionless spatial coordinate $x$ is measured in the units
$L/\pi$ where $L$ is the lattice period in the physical units. Thus,
both lattices, i.e. both functions $V(x)$ and $\Omega_{3}(x)$, are
$\pi$-periodic, $h_{\epsilon}(x)=h_{\epsilon}(x+\pi)$. We also assume
that the lattices are even with respect to the origin, i.e, $h_{\epsilon}(x)=h_{\epsilon}(-x)$,
and odd with respect to the quarter-period, i.e., $h_{\epsilon}(\pi/4+x)=-h_{\epsilon}(\pi/4-x)$.
For instance, one can consider $\{V,\Omega_{3}\}\propto\cos(2x)$.
In the units, where $\hbar=1$ and $m=1/2$, the evolution of the
spinor $|\Psi\rangle$ is described by the Schr\"odinger-Pauli equation:
\begin{eqnarray}
i\frac{\partial|\Psi\rangle}{\partial t}=H|\Psi\rangle+\beta(t)x|\Psi\rangle,\quad H\equiv H_{0}+h_{\epsilon}(x),
\label{SE}
\end{eqnarray}
where
\begin{eqnarray}
H_{0}=p^{2}+\gamma\sigma_{3}p+\Omega_{1}\sigma_{1},\label{H0}
\end{eqnarray}
$p=-i{\partial_{x}}$ is the operator of linear momentum, $\gamma$
is the strength of the SO coupling, and $2\Omega_{1}$ is the Rabi
frequency (further we set $\Omega_{1}=0.5$). The term $\beta(t)x$
takes into account the linear force oscillating with the frequency
$\omega$ and having zero mean value on one oscillation period.

We are interested in the evolution of the wavepacket center which
can be characterized by its coordinate $x_{c}$ defined as the expectation
value
\begin{eqnarray}
x_{c}=\langle\Psi|x|\Psi\rangle=\int_{-\infty}^{\infty}\bPsi^{\dag}x\bPsi dx.\label{average}
\end{eqnarray}

Below we use the complete set of Bloch states $|k,\alpha\rangle$
of the Hamiltonian $H$: $H|k,\alpha\rangle=\mu_{\alpha}(k)|k,\alpha\rangle$,
where $\mu_{\alpha}(k)$ is the energy of the state, the Bloch quasi-momentum
$k$ belongs to the reduced Brillouin zone (BZ): $k\in(-1,1]$, and
$\alpha=1,2,...$ is the band number ($\alpha=1$ being the number
of the lowest band). Following Houston's approach~\cite{Houston},
we introduce adiabatically-varying Bloch states $|\kappa (t),\alpha\rangle$
where
\begin{eqnarray}
\kappa (t)=k+B(t),\qquad B(t)=\int_{0}^{t}\beta(t')dt',\label{B(t)}
\end{eqnarray}
and $k$ is the initial value of the quasi-momentum (or central
quasi-momentum in the case of localized Bloch wavepackets). The spinor
$|\Psi\rangle$ can be expanded in terms of the adiabatically-varying states as:
\begin{eqnarray}
|\Psi\rangle=\sum_{\alpha=1}^{\infty}\int_{-1}^{1}dk\chi_{\alpha}(k,t)|\kappa(t),\alpha\rangle.\label{expan_new}
\end{eqnarray}
and, for the sake of convenience, the spectral coefficients will be
represented as
\begin{eqnarray}
\chi_{\alpha}(k,t)=\chi_{\alpha}^{(0)}(k,t)e^{-i\int_{0}^{t}\mu_{\alpha}[\kappa(\tau)]d\tau}\label{chi_1}
\end{eqnarray}
with the functions $\chi_{\alpha}^{(0)}(k,t)$ to be determined latter.
 The normalization condition (in the direct and Fourier spaces)
\begin{eqnarray}
\langle\Psi|\Psi\rangle=\sum_{\alpha=1}^{\infty}\int_{-1}^{1}|\chi_{\alpha}(k,t)|^{2}dk=1\label{normal}
\end{eqnarray}
is also imposed.

We assume that the following conditions hold:

\paragraph{Condition 1:}

Bloch states of only one band, say of the band $\alpha_{0}$, are
initially excited, i.e. $\chi_{\alpha}(k,0)=0$ for $\alpha\neq\alpha_{0}$.

\paragraph{Condition 2:}

A wavepacket $|\Psi\rangle$ is a Bloch wave with a smooth and broad envelope
(compared to the lattice period) and its spectrum centered
at a quasi-momentum $k_{0}$ in the reduced BZ is much
narrower than the BZ zone, so that the approximation
\begin{eqnarray}
\int_{-1}^{1}dk|\chi_{\alpha_0}(k,t)|^{2}\frac{\partial\mu_{\alpha_0}(k)}{\partial k}\approx\frac{\partial\mu_{\alpha}(k_{0})}{\partial k_{0}}
\label{eq:approx_mu}
\end{eqnarray}
is valid {[}here we take into account the normalization \eqref{normal}{]}.

\paragraph{Condition 3:}

The linear force is weak, i.e., $|\beta(t)|\ll1$, so that inter-band
tunneling is negligible.

Under these conditions one can show that the the dynamics of the wavepacket
center is governed by the equation  (see \cite{KKZT} for the derivation):
\begin{eqnarray}
\frac{dx_{c}}{dt}=\left.\frac{\partial\mu_{\alpha_{0}}(k)}{\partial k}\right|_{k=k_{0}+B(t)},\label{xc_dyn}
\end{eqnarray}
where $k_{0}$ is the initial central quasi-momentum of the wavepacket.

Taking into account that due to  Floquet theorem the energy is
a periodic function, with period 2, $\mu_{\alpha_{0}}(k)$ can
be expanded in a Fourier series. Generally, if the excited band
$\alpha_{0}$ is narrow enough, one can restrict the consideration
to only one harmonic associated with nonzero frequency --- this is a
standard assumption in the tight-binding approximation. However, in
our system with SO coupling this is usually not the case (see Fig.~\ref{fig:one}
with representative band structures), and one has to take into account
several Fourier harmonics in the expansion
\begin{eqnarray}
\label{eq:omit}
\mu_{\alpha}(k)=\sum_{m=0}^{\infty}\mu_{\alpha}^{(m)}\cos(m\pi k).\label{Fourier_mu}
\end{eqnarray}
(In order to simplify   analytical expressions, in (\ref{eq:omit}) as well as in the subsequent formulas we omit subscript $0$ at $\alpha$ and write $\alpha$ instead of $\alpha_0$  bearing in mind that only one zone is being considered.)
This yields the final form of the equation describing the dynamics
of the center of the wavepacket:
\begin{eqnarray}
\frac{dx_{c}}{dt}=-\pi\sum_{m=1}^{\infty}m\mu_{\alpha}^{(m)}\sin\{m\pi[k_{0}+B(t)]\}.
\label{eq:dyn_final}
\end{eqnarray}

Now we consider a periodic force $B(t)=B(t+T)$, with the period $T$
and frequency $\omega=2\pi/T$. Dynamic localization is characterized by the vanishing average velocity of the wavepacket,
where averaging is performed over the period $T$, as well as by 
restoration of the wavepacket shape after each period, i.e., by suppression
of dispersion. Thus the resonant frequency $\omega$ at which the
dynamic localization occurs [i.e. at which $x_{c}(0)=x_{c}(T)$] is determined
from the condition
\begin{equation}
\sum_{m=1}^{\infty}m\mu_{\alpha}^{(m)}\int_{0}^{T}\sin\{m\pi[k_{0}+B(t)]\}dt=0.\label{average_omega}
\end{equation}

First, we consider a general harmonic force of the form
$\beta(t)=\beta_{0}\sin(\omega t+\varphi)$
where $\beta_{0}$ and $\varphi$ are the amplitude
and the initial phase of the force. Now the integrals in (\ref{average_omega})
are computed explicitly resulting in the condition 
\begin{eqnarray}
\label{average_final}
{\cal J}(\omega,k_{0})\equiv\sum_{m=1}^{\infty}m\mu_{\alpha}^{(m)}\sin\left[\pi m\left(k_{0}+\frac{\beta_{0}}{\omega}\cos\varphi\right)\right]\nonumber \\
\times J_{0}\left(\frac{m\pi\beta_{0}}{\omega}\right)=0,\quad
\end{eqnarray}
where $J_{0}(\cdot)$ is the zero-order Bessel function of the first
kind. Equation (\ref{average_final}) will
be used below for the comparison of analytical predictions with the
results of direct numerical simulations. \\

An immediate outcome of (\ref{average_final}) follows from the fact
that 
$\sin(m\nu)=\sin\nu U_{m-1}(\cos\nu)$,
where $U_{m}(\cdot)$ is the Chebyshev polynomial of the second kind:
independently of the type of the lattice, i.e. independently of $\epsilon$
in   Eq.~\eqref{HL}, the sum in Eq.~(\ref{average_final}) is
proportional to
\begin{eqnarray}
\sin\left[\pi\left(k_{0}+\frac{\beta_{0}}{\omega}\cos\varphi\right)\right].
\end{eqnarray}
Thus the function ${\cal J}(\omega,k_{0})$ vanishes at the frequencies
$\omega=\tom_{n}$ defined as
\begin{equation}
\tom_{n}=\frac{\beta_{0}\cos\varphi}{n-k_{0}},\label{pseudo_zero}
\end{equation}
where $n$ is an integer. These zeros, however do not represent {\em
authentic resonances} corresponding to the dynamic localization,
but reflect the adiabatic theorem~\cite{adiabatic} ensuring variation
of the quasi-momentum caused by slowly varying weak external force.
The zeros $\tom_{n}$ depend on the central momentum $k_{0}$ (as
well as on the initial phase of the force $\varphi$) and thus at
$\omega=\tom_{n}$ the dynamics is accompanied by the {\em effective
dispersion} of the wavepacket. To separate the resonances at $\tom_{n}$
from the {\em authentic resonances} $\omega_{k}$ leading to the dynamic
localization, the former ones will be termed {\em pseudo-resonances}.
When only one harmonic ($m=1$) is taken into account
in the expansion of the band energy into Fourier series, authentic
resonances can be found using zeros of the Bessel functions in Eq.
(\ref{average_final}). In this case $\omega_{k}$ are approximately
given by $\omega_{k}=\pi\beta_{0}/\nu_{k}$, where $\nu_{1}\approx2.405$,
$\nu_{2}\approx5.520$, etc., are zeros of $J_{0}(\nu)$. In the authentic
resonances computed within this tight-binding approximation the dispersion
usually vanishes too and the wavepacket undergoes periodic in time
localization.

We concentrate on the case $\varphi=0$, i.e.,
\begin{equation}
\beta(t)=\beta_{0}\sin(\omega t),\label{eq:harm}
\end{equation}
and consider initial wavepackets with $k_{0}=0$ exciting modes in
the center of the first BZ. Due to relative smoothness of the band
structure (see Fig.~\ref{fig:one} below), for accurate description
of all resonances it is usually enough to take into account only a
few harmonics in (\ref{average_final}) (see Tables~\ref{tab:one}
and ~\ref{tab:two} below). Thus, the resonant frequencies
at which dynamic localization occurs are found from the equation
\begin{eqnarray}
\label{average_M}
{\cal J}_{M}(\omega)\equiv\mu_{\alpha}^{(1)}J_{0}\left(\frac{\pi\beta_{0}}{\omega}\right) \quad\quad\quad\quad\quad\quad\quad\quad\quad\nonumber \\
+\sum_{m=2}^{M}m\mu_{\alpha}^{(m)}\frac{\sin\left(m\pi\beta_{0}/\omega\right)}{\sin\left(\pi\beta_{0}/\omega\right)}J_{0}\left(\frac{m\pi\beta_{0}}{\omega}\right).%
\end{eqnarray}

The expressions (\ref{average_M}) {[}or (\ref{average_final}){]} reveal
a noticeable result: the more complex the dependence $\mu_{\alpha}(k)$,
i.e. the larger the number of Fourier harmonics needed for its
accurate approximation, the more pronounced is the {\em effective
dispersion}. In order to estimate the variation of the width of the wavepacket,
we assume that initially it has the width $d_{0}$, and hence
its spectral width can be defined as $\delta k=2\pi/d_{0}$. Since
resonant frequencies depend on the quasi-momentum, one can estimate the
difference between resonant periods of the linear force for excitations
separated by $\delta k$ in the frequency domain as:
\begin{eqnarray}
\delta T\approx-\frac{2\pi}{\omega^{2}}\left(\frac{d\omega}{dk}\right)_{k_{0}}\!\!\delta k=\frac{4\pi^{2}}{\omega^{2}d_{0}}\frac{\partial{\cal J}(\omega,k_{0})/\partial k_{0}}{\partial{\cal J}(\omega,k_{0})/\partial\omega},\label{estim_T}
\end{eqnarray}
where ${\cal J}(\omega,k)$ was defined in (\ref{average_final})
and the derivative of $d\omega(k)/dk$ was computed using the
implicit relation ${\cal J}(\omega,k)=0$. Respectively, the variation
of the wavepacket width over one period of the linear force can be estimated
as the absolute value of the product of the difference of the resonant
periods $\delta T$ for two harmonics separated by $\delta k$ and
the difference of their group velocities:
\[
{\displaystyle {\left(\frac{d\mu_{\alpha}(k)}{dk}\right)_{k_{0}}-\left(\frac{d\mu_{\alpha}(k)}{dk}\right)_{k_{0}+\delta k}}.}
\]
Since in our case $k_{0}=0$ and $d\mu_{\alpha}(k)/dk|_{k=0}=0$ the
required estimate reads
\begin{eqnarray}
\delta d\approx\delta T\cdot\left(\frac{d\mu_{\alpha}(k)}{dk}\right)_{\delta k}\approx\frac{(2\pi)^{3}\mu_{\alpha}^{(1)}}{\omega^{2}d_{0}^{2}}\left(\frac{{\cal J}_{k}(\omega,k)}{{\cal J}_{\omega}(\omega,k)}\right)_{k=0}. \label{estim_width}
\end{eqnarray}
Here we took into account only the first harmonic in the expansion
(\ref{Fourier_mu}) of $\mu_{\alpha}(k)$ in the Fourier series.

In the tight-binding approximation, where only one Fourier harmonic
is taken into account, i.e. $M=1$, one readily gets $\left({d\omega}/{dk}\right)_{k=k_{0}}=0$,
i.e. $\delta d=0$ as described in many previous studies.

\section{Dynamic localization in optical lattices}

\label{sec:OL}

\begin{figure}
\begin{centering}
\includegraphics[width=1\columnwidth]{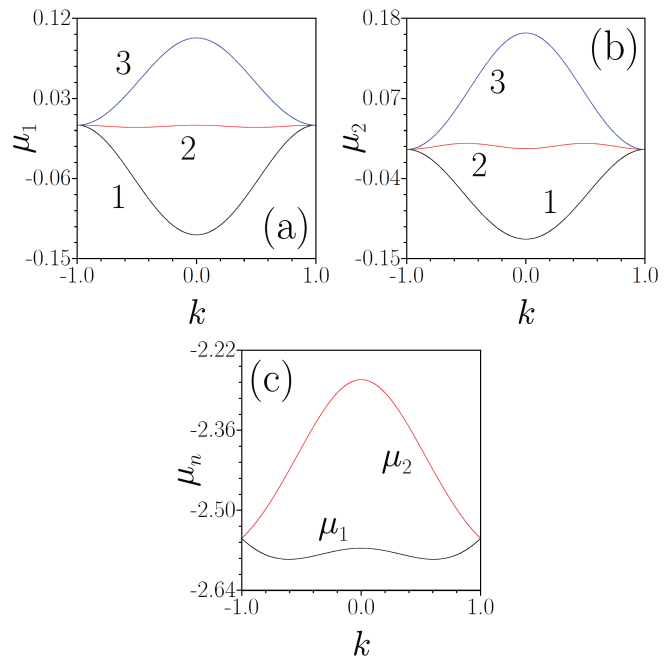}
\par\end{centering}

\caption{(Color online) Transformation of the first (a) and second (b) bands
in the spectrum for an OL with growing $\gamma$. In (a) $\gamma=0$ for
curve 1, $\gamma=1.17$ for curve 2, and $\gamma=2.15$ for curve
3. In (b) $\gamma=0$ for curve 1, $\gamma=0.82$ for curve 2, and
$\gamma=1.84$ for curve 3. The vertical shift of the bands with increasing $\gamma$ were eliminated by subtracting $\mu_{n}(k=1)$ from the $\mu_{n}(k)$
dependence. (c) First two bands in the spectrum for a ZL at $\gamma=2$. }
\label{fig:one}
\end{figure}

\begin{figure}
\begin{centering}
\includegraphics[width=1\columnwidth]{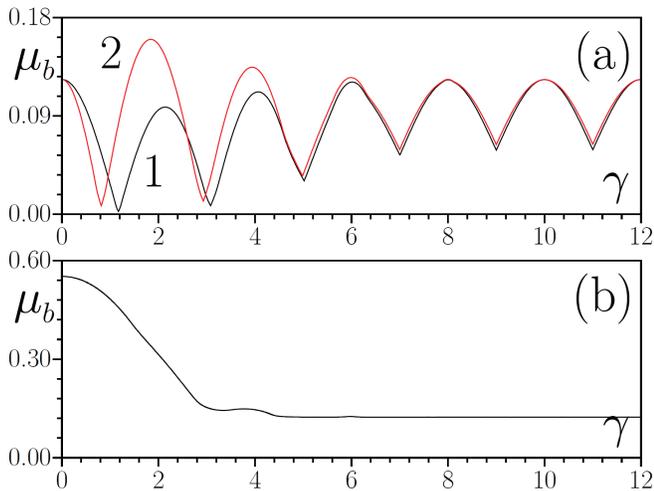}
\par\end{centering}

\caption{(Color online) (a) Width of the first (curve 1) and second (curve
2) bands in an OL {\em vs} $\gamma$. (b) Width of the combined
band in a ZL {\em vs} $\gamma$. }
\label{fig:two}
\end{figure}

First we consider an atom in an OL 
where the Hamiltonian in Eq.~(\ref{SE}) takes the form $H=H_{0}+h_{1}(x)=H_{0}+V(x)I$.
In all numerical simulations we consider $V(x)\equiv-4\cos(2x)$ and
concentrate on an atom whose energy belongs to one of the two lowest
bands, shown in Fig.~\ref{fig:one}(a) and Fig.~\ref{fig:one}(b)
for several values of the SO coupling. In these panels one observes
well-known effects~\cite{KKZT,YoChu} such as collapse of the band
width (red curves 2) and inversion of the curvature (seen from the
comparison of the curves 1 and 3) upon increase of the strength of the
SO coupling. It is this effect that drastically affects the phenomenon
of dynamic localization in SO-coupled systems, since according to
Eq. (\ref{xc_dyn}) the amplitude of oscillations of the wavepacket
under the action of a periodic linear force is determined by the width
of the excited band. The widths of the first two bands in the spectrum
are shown in Fig.~\ref{fig:two}(a) as functions of the SO-coupling
strength $\gamma$. Notice that both first and second bands nearly
collapse for a pair of $\gamma$ values, but these pairs are different
for the first and second bands. For large values of $\gamma$ the
width of the band oscillates nearly periodically.

\begin{figure}
\includegraphics[width=1\columnwidth]{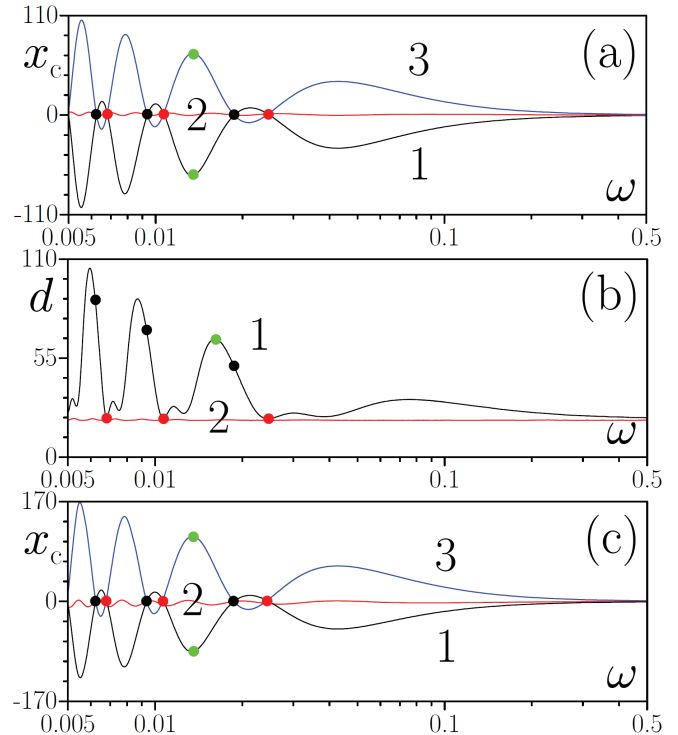} \caption{(Color online) Resonant curves showing (a),(c) position of the integral
center of the wavepacket and (b) its width at $t=6\pi/\omega$ {\em
vs} modulation frequency $\omega$ in the OL. Panels (a),(b) correspond
to the first-band excitation at $\gamma=0.5$ (curve 1), $\gamma=1.17$
(curve 2), and $\gamma=2.15$ (curve 3). Panel (c) corresponds to
the second-band excitation at $\gamma=0$ (curve 1), $\gamma=0.82$
(curve 2), and $\gamma=1.84$ (curve 3). Red dots show resonances
where there is no drift and broadening of the wavepacket, while black
dots show resonances where wavepacket broadens, but does not drift.
Green dots mark frequencies at which largest displacement or broadening
occurs between two first groups of resonances. In all cases $d(t=0)=7\pi$.
Notice the logarithmic scale in the horizontal axis. }
\label{fig:three}
\end{figure}

Now we turn to the numerical simulations of the dynamic localization. To this end, we focus on the particular amplitude of
the linear force $\beta_{0}=1/(17\pi)\approx0.019$. In Fig.~\ref{fig:three}
we illustrate the main result obtained for an OL in the form of resonant
curves showing dependence of the shift of the wavepacket center $x_{c}$
and its integral width $d$ on the frequency of linear force calculated
after three periods of the force oscillations, $t=3T=6\pi/\omega$.
Here and in all simulations of evolution dynamics we use the initial
wavepacket in the form of modulated Bloch wave from the center of
the BZ, i.e.,
\[
|\Psi(t=0)\rangle=e^{-(2x/d_{0})^{2}}|0,\alpha\rangle,
\]
where 
$\alpha=1$
in panels (a),(b) and $\alpha=2$ in panel (c). The instantaneous
width $d(t)$ of the wavepacket was computed using the inverse form-factor,
i.e. as
\[
d(t)=\left[\int_{-\infty}^{\infty}(|\psi_{1}|^{4}+|\psi_{2}|^{4})\,dx\right]^{-1},
\]
i.e. initially $d(0)=d_{0}$. 

With red and black dots in Fig.~\ref{fig:three}(a) we show resonances
and pseudo-resonances that for the first band can be clearly identified
by comparing panels (a) and (b): while in the authentic resonances
(red dots) we observe no displacement of the center of the wavepacket
and complete restoration of its input width after three periods, in
pseudo-resonances (black dots) we still do not observe displacement,
but dispersion can be rather strong. Resonances and
pseudo-resonances alternate: between two resonances one always encounter
one pseudo-resonance. Importantly, due to the collapse of the first band
at $\gamma\approx1.17$ and of the second band at $\gamma\approx0.82$
{[}red curves 2 in Fig.~\ref{fig:three}(a) and (b), respectively{]},
the oscillations of excitations from the corresponding bands are strongly
suppressed, as well as their dispersion. In this regime (i.e., at
these specific values of $\gamma$) the wavepacket propagates without
broadening and displacement for any frequency $\omega$ of the linear
force, so that dynamic localization does not occur. Notice
also that the direction of the wavepacket displacement changes when the
SO-coupling strength $\gamma$ passes the critical value corresponding
to the band collapse {[}compare lines 1 and 3 in Fig.~\ref{fig:three}(a){]}.
Therefore, in this system one can control the direction of the wavepacket
average displacement by changing the SO-coupling strength. The structure
of resonances is similar for the first- and second-band excitations
[compare panels Fig.~\ref{fig:three}(a) and \ref{fig:three}(c)].

The resonant and pseudo-resonant frequencies obtained numerically
agree remarkably well with the predictions based on the expression (\ref{average_M})
even when only two harmonics in the Fourier expansion (i.e., $M=2$)
are used. This is illustrated in Table~\ref{tab:one} for several
values of the SO coupling. It is relevant to mention that in spite
of the smallness of the second harmonic ($|\mu_{\alpha}^{(2)}|\ll|\mu_{\alpha}^{(1)}|$), taking it into account
does improve the accuracy of the calculation of the resonant frequency.
For example, when the first two resonance frequencies $\omega_{num}=0.02472$
and $\omega_{num}=0.0107$ at $\gamma=0$ are calculated using only
one harmonic in the Fourier expansion (tight-binding approximation)
one obtains $\omega_{1harm}=0.02445$ and $\omega_{1harm}=0.01065$,
respectively.

\begin{table}
\begin{tabular}{c | c|c|c|}
	\hline
	\multicolumn{1}{|c|}{\multirow{1}{*}{$\gamma$} }  & band & Fourier coefficients & $\omega_{2harm}$ \quad $\omega_{num}$
	\\
	\cline{1-4}
	\multicolumn{1}{ |c  }{\multirow{2}{*}{\begin{tabular}{l}
					\\ 	\\ 0 	\\ 	
				\end{tabular}} } &
				\multicolumn{1}{ |c| }{1} & \begin{tabular}{l}
					$\mu_1^{(1)}\approx  -0.0616$ 	\\ 	$\mu_1^{(2)}\approx-0.00127$ 	
				\end{tabular} &
			 	\begin{tabular}{cc}
					0.02476 & 0.02472
					\\
			\cellcolor[gray]{0.9}	 0.01872 & 0.01873
					\\
					 0.01071 & 0.01070
					\\
			\cellcolor[gray]{0.9} 0.00936 &  0.00936
					\\
				0.00682	& 0.00682
    			\end{tabular}
	\\ \cline{2-4}
				\multicolumn{1}{ |c  }{}                        &
				\multicolumn{1}{ |c| }{2} &\begin{tabular}{l}
					$\mu_2^{(1)}\approx  -0.06160$
					\\
					$\mu_2^{(2)}\approx  0.00127$
 				\end{tabular}
				&
				\begin{tabular}{cc}
					 0.02476 & 0.02472
					\\
			\cellcolor[gray]{0.9}  0.01872 & 0.01873
					\\
					0.01071 & 0.01070
					\\
			\cellcolor[gray]{0.9} 0.00936 & 0.00936
				\\
			0.00682	& 0.00682
				\end{tabular}
				\\ \cline{1-4}
\multicolumn{1}{ |c  }{\multirow{2}{*}{\begin{tabular}{l}
			\\ 	\\ 0.5 	\\ 	
		\end{tabular}} } &
						\multicolumn{1}{ |c| }{1} & \begin{tabular}{l}
							$\mu_1^{(1)}\approx -0.04802$ 	\\ 	$\mu_1^{(2)}\approx 0.00134$
						\end{tabular} &
						\begin{tabular}{cc}
					0.02487 &  0.0248
							\\
			\cellcolor[gray]{0.9}  0.01872	  & 0.01870
							\\
					0.01073	& 0.01072
							\\
			\cellcolor[gray]{0.9}  0.00936 &  0.00936
							\\
						0.00683	& 0.00682
						\end{tabular}
						\\ \cline{2-4}
						\multicolumn{1}{ |c  }{}        &
						\multicolumn{1}{ |c| }{2} &\begin{tabular}{l}
							$\mu_2^{(1)}\approx -0.03544$
							\\
							$\mu_2^{(2)}\approx  -0.00172$
						\end{tabular}
						&
						\begin{tabular}{cc}
						0.02387	  & 0.02472
							\\
			\cellcolor[gray]{0.9}  0.01872 & 0.01872
							\\
						0.01054	 & 0.01056
							\\
			\cellcolor[gray]{0.9} 0.00936 & 0.00937
						\\
					0.00675	& 	0.00675
						\end{tabular}
						\\ \cline{1-4}
						 \multicolumn{1}{ |c  }{\multirow{2}{*}{\begin{tabular}{l}
											\\ 	\\  2.5	\\ 	
										\end{tabular}} } &
										\multicolumn{1}{ |c| }{1} & \begin{tabular}{l}
											$\mu_1^{(1)}\approx 0.04067$ 	\\ 	$\mu_1^{(2)}\approx 0.00183$
										\end{tabular} &
										\begin{tabular}{cc}
										0.02391	& 0.02390
											\\
						\cellcolor[gray]{0.9}  0.01872 & 0.01870
											\\
										0.01055	& 0.01055
											\\
						\cellcolor[gray]{0.9} 0.00936 &  0.00936
										\\
								0.00675		& 0.00675
										\end{tabular}
										\\ \cline{2-4}
										\multicolumn{1}{ |c  }{}        &
										\multicolumn{1}{ |c| }{2} &\begin{tabular}{l}
											$\mu_2^{(1)}\approx  0.04541$ \\ 	$\mu_2^{(2)}\approx -0.00182$
										\end{tabular}
										&
										\begin{tabular}{cc}
								0.02506		  & 0.02501
											\\
            			\cellcolor[gray]{0.9} 0.01872		 &  0.01872
											\\
										0.01077	 & 0.01076
											\\
			\cellcolor[gray]{0.9} 0.00936  &  0.00936
										\\
									0.00684	& 0.00684
						\end{tabular}
						\\ \cline{1-4}
					\end{tabular}
					\caption{ Coefficients of the Fourier expansion of the $\mu_\alpha(k)$ dependence; resonance and pseudo-resonance (in shadowed cells) frequencies $\omega_{2harm}$ computed using  two Fourier harmonics in the expansion (the coefficient $\mu_0$ does not contribute to   dynamics),
					and   resonant and pseudo-resonant frequencies  $\omega_{num}$ obtained from the direct numerical simulations (see Fig.~\ref{fig:three}). }
						\label{tab:one}
				\end{table}

\begin{figure*}[t]
\begin{centering}
\includegraphics[width=1\textwidth]{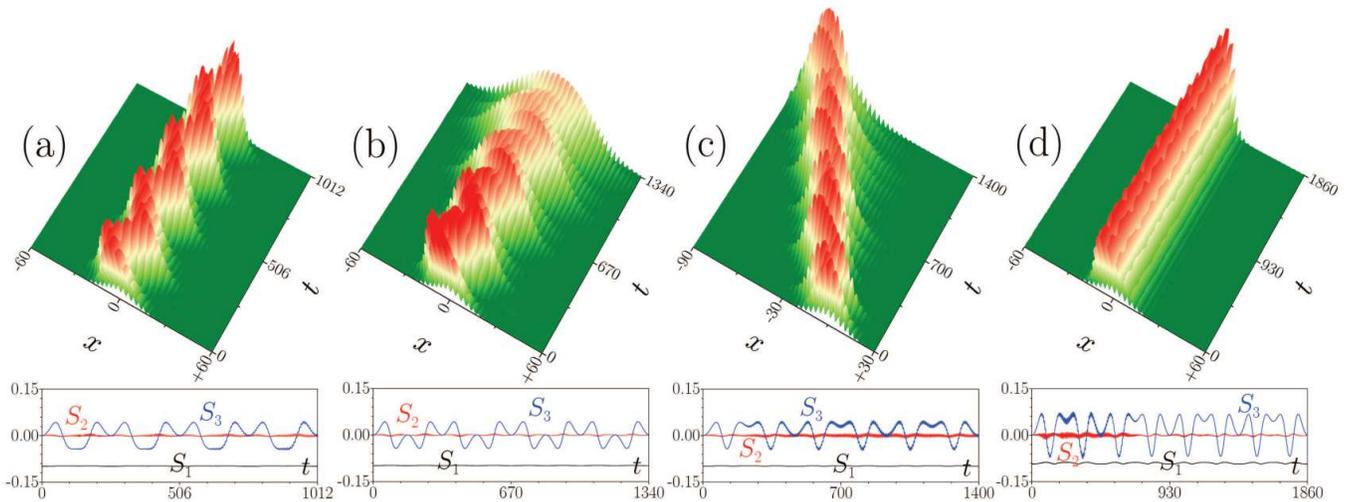}
\par\end{centering}

\caption{(Color online) Evolution dynamics for the first-band excitations in
an OL at (a) $\omega=0.0248$, $\gamma=0.5$, (b) $\omega=0.0187$,
$\gamma=0.5$, (c) $\omega=0.0135$, $\gamma=0.5$, and (d) $\omega=0.0135$,
$\gamma=1.17$. Only $|\psi_{1}|$ component is shown. The second row
shows the corresponding evolution of the spin components. For illustrative
purposes the magnitude of the $S_{1}$-component was divided by five.
In all cases $d(t=0)=7\pi$. In panels (a), (b), and (d) the evolution
time corresponds to four complete oscillation periods, while in (c)
it corresponds to three oscillation periods.}
\label{fig:four}
\end{figure*}

Different evolution scenarios of a Bloch wavepacket from the first
band are shown in Fig.~\ref{fig:four} for different values of the frequencies
of the linear force. Taking into account the spinor nature of the
wavefunction we show both the amplitude of the wavepacket (the upper
row) and the expectation values for the spin components $S_{j}=\langle\Psi|\frac{1}{2}\sigma_{j}|\Psi\rangle$
(the lower row). Panels (a) and (b) show the evolution in the first authentic
resonance and pseudo-resonance at $\gamma=0.5$ (see also Fig.~\ref{fig:three}(a),
(b) and Table~\ref{tab:one}). As mentioned above, in an authentic
resonance the wavepacket experiences neither drift nor broadening,
while in a pseudo-resonance there is no drift, but one can observe
considerable expansion. The drift of the wavepacket accompanied by slow
dispersion at a non-resonant frequency is shown in panel (c), while
complete suppression of oscillations and dispersion (even for a non-resonant
frequency) in the case where band collapse occurs is illustrated in
panel (d).

In all cases the motion of the Bloch wavepacket is accompanied
by the spin-wave. A common property of all four panels in the lower
row in Fig.~\ref{fig:four} is that the spin slightly oscillates
in the $(x,z)$-plane with almost constant $x$-component and negligible
$y$-component. This can be understood from the system of equations
for the expectations of the spin components that
admits the exact relation ${dS_{3}}/{dt}=2\Omega_{1}S_{2}$, which
for $S_{2}(t=0)=0$ leads to $S_{2}/S_{3}\sim1/(\Omega_{1}T)=\omega/(2\pi\Omega_{1})$.
Thus the relation between the $y-$ and $z-$components is determined
by the relation between the frequency of the linear force and the Rabi
frequency. In the examples shown in Fig.~\ref{fig:four}, $\omega\sim10^{-2}$
while $\Omega_{1}=0.5$. This leads to the estimate $S_{2}/S_{3}\sim10^{-3}$
consistent with the negligible $y$-projection of the spin. Notice that the
evolution of the $S_{3}$ component in ``spin-up'' $S_{3}>0$ and
``spin-down'' $S_{3}<0$ half-periods is different in all plots
in Fig.~\ref{fig:four}, as a consequence of the broken parity symmetry
in the $z$-direction due to the presence of the SO coupling.

\begin{figure}
\begin{centering}
\includegraphics[width=1\columnwidth]{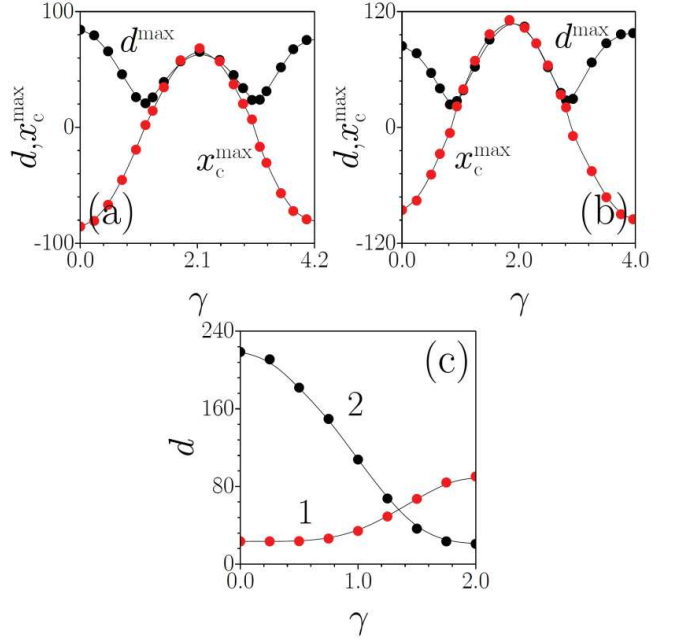}
\par\end{centering}

\caption{(Color online) Maximal width of the wavepacket $d^{{\rm max}}$ and
its maximal displacement $x_{c}^{{\rm max}}$ between the two highest
authentic resonances at $t=6\pi/\omega$ {\em vs} $\gamma$ for
the first-band (a) and second-band (b) excitations in the OL. In all
cases $d(t=0)=7\pi$. (c) Width of the wavepacket at $t=6\pi/\omega$
in the first (curve 1) and second (curve 2) resonances vs. $\gamma$
for the first-band excitation in the ZL. In all cases $d(t=0)=7\pi$. }
\label{fig:five}
\end{figure}

In Fig.~\ref{fig:five}(a) and (b) we show the maximal displacement
$x_{c}^{{\rm max}}$ (i.e. $|x_{c}^{{\rm max}}|$ is the amplitude
of oscillations of the wavepacket center) and the maximal width of
the wavepacket $d^{{\rm max}}$ achieved in the frequency interval
between the first two authentic resonances, as the functions of the SO
coupling. The values $x_{c}^{{\rm max}}$ and $d^{{\rm max}}$ were
calculated numerically for excitations emerging from the first [Fig.~\ref{fig:five}(a)]
and second {[}Fig.~\ref{fig:five}(b){]} bands at $t=6\pi/\omega$
(after three periods of the linear force). In Fig.~\ref{fig:three}
the maximal width of the wavepacket achieved are shown
by green dots in the respective panels. Notice that the maximal amplitude
and the maximal width are achieved at slightly different frequencies
(visible in the logarithmic scale of Fig.~\ref{fig:three}). The
dependencies shown in Fig.~\ref{fig:five}(a) and \ref{fig:five}(b)
qualitatively reproduce band width from Fig.~\ref{fig:two}(a): the larger band width results
in a larger amplitude of oscillations and larger dispersion of the wavepacket,
while narrow band strongly suppresses both processes.  {\color{black} We also observe that the closer the modulation frequency to one of the resonances, the smaller the smearing out of the probability density pattern due to dispersion. This implies that small deviations of modulation frequency from its resonant value do  not result in a dramatic change of the dynamics.}

\begin{figure}
	\begin{centering}
		\includegraphics[width=1\columnwidth]{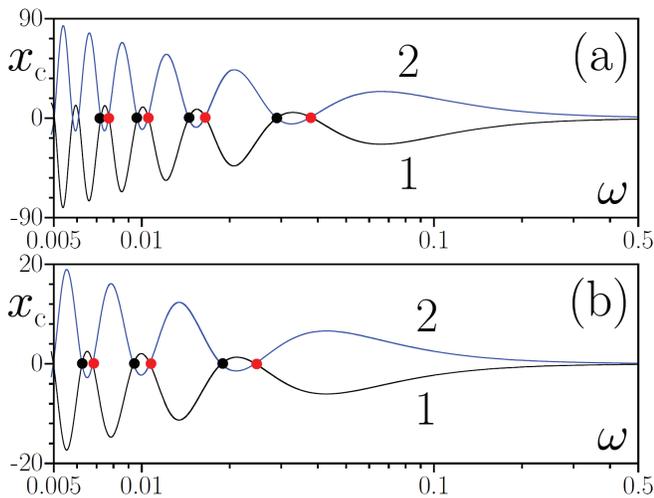}
		\par\end{centering}
	
	\caption{{\color{black}(Color online) Resonant curves showing position of the integral center of broad wavepacket from the first band versus modulation frequency   in the OL at $\gamma=0.5$ (curve 1) and $\gamma=2.15$ (curve 2). In (a) $\beta=1/11\pi$ and lattice depth is $4$, while in (b) $\beta=1/17\pi$  and lattice depth is $8$. Red dots -- authentic resonances, black dots -- pseudo-resonances.} }
	\label{fig:five_a}
\end{figure}

{\color{black} To end this section, we recall that while the analytical results are general, all numerical simulations presented here were performed for specific values of the strength of the linear force and of the potential depth. However, we have verified that changing the values of such parameters  does not introduce qualitative differences in the obtained results, as illustrated  in Fig.~\ref{fig:five_a}.  All resonant curves are qualitatively similar to the curves obtained for other parameters, as shown in Fig.~\ref{fig:three}. As it is expected from Eq.~(\ref{average_final}) [or from Eq.~(\ref{average_M})] in Fig.~\ref{fig:five_a} (a) one observes a shift of the resonances and pseudo-resonances towards larger frequencies as $\beta$ increases. Fig.~\ref{fig:five_a} (b) shows that a change of the potential depth does not affect the location of the resonant frequencies, but  it does significantly reduce the amplitude of the resonant curves [cf. Fig.~\ref{fig:three} (a)]. This is in full agreement with the decrease of the band widths with increasing the potential depth. Therefore, one concludes that dynamic localization can be observed for a large range of lattice depths and linear forces, by tuning the frequency of modulation.}

\section{Dynamic localization in Zeeman lattices}

\label{sec:ZL}

\begin{figure}
\begin{centering}
\includegraphics[width=1\columnwidth]{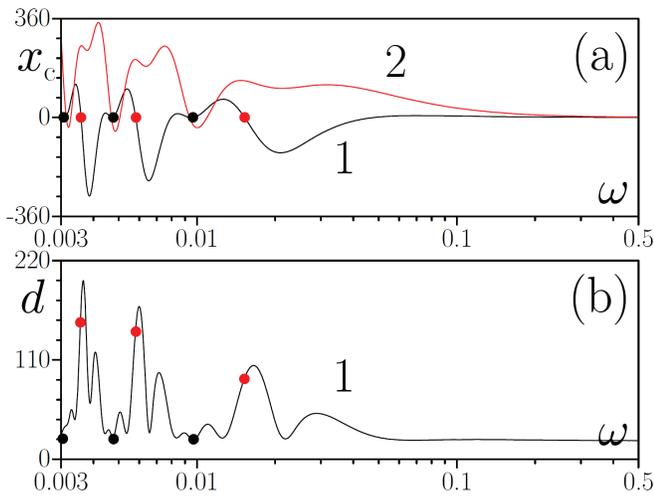}
\par\end{centering}

\caption{(Color online) Resonant curves showing (a) position of the average
center of the wavepacket and (b) its width at $t=6\pi/\omega$ {\em
vs} modulation frequency $\omega$ at $\gamma=2$ in the ZL. First-band
(curve 1) and second-band (curve 2) excitations are shown in (a),
while in (b) only first-band excitation is shown. Dots indicate resonances
for the first-band excitation, which differ from resonances for second-band
excitation. In all cases $d(t=0)=7\pi$. The largest resonance $0.04822$
(see Table~\ref{tab:two}) is not shown.}
\label{fig:six}
\end{figure}

Now we consider the phenomenon of dynamic localization in a ZL.
While formally the formulas derived in Sec.~\ref{sec:general} remain
valid (they were derived without any assumptions about the type of
lattice potential), their physical interpretation has to be revised. The motivation is twofold:
the wavepacket is two-component (spinor) and the bands of the ZL cross each other
at the edges of the BZ {[}see Fig.~\ref{fig:one}(c){]}.
It was shown in \cite{KKZT} that when the wavepacket moves adiabatically
along the selected branch of the band and arrives at the opposite
edge of the BZ the associated spinor flips. In other words, unlike
in an OL, in the case of ZLs, after crossing the BZ a spinor
does not recover its initial state. Moreover it does not continue
moving in the initial band, but passes to the next band with which
the initial band crosses at $k=\pm1$. Let us assume that the indexes
of the crossing bands are $\alpha$ and $\alpha+1$ {[}in Fig.~\ref{fig:one}
these are bands 1 and 2, i.e., $\alpha=1${]}. Suppose also that the
initial wavepacket belongs to the $\alpha$-th band. Then, after the second
crossing of the BZ but now in the band $\alpha+1$, the spin undergoes
a new flip and the spinor returns to its original state. Hence, to
compute the time of restoration of the wavepacket one has to consider the
double crossing of the {\em reduced} BZ. Alternatively,
and technically more convenient, one can consider the adiabatic motion
in the {\em extended\/} BZ, which in our case is $k\in(-2,2]$.
Then, instead of considering $\mu_{\alpha}(k)$ and $\mu_{\alpha+1}(k)$
in the reduced BZ, one considers
\begin{subequations}
\label{eq:mu_til}
\begin{eqnarray}
\tmu_{\alpha}=\left\{ \begin{array}{ll}
\mu_{\alpha}(k), & k\in(-1,1]\\
\mu_{\alpha+1}(k), & k\in(-2,-1]\cup(1,2]
\end{array}\right.
\end{eqnarray}
and
\begin{eqnarray}
\tmu_{\alpha+1}=\left\{ \begin{array}{ll}
\mu_{\alpha+1}(k), & k\in(-1,1]\\
\mu_{\alpha}(k), & k\in(-2,-1]\cup(1,2].
\end{array}\right.
\end{eqnarray}
\end{subequations}
Thus, $\tmu_{\alpha}(k)=\mu_{\alpha+1}(k\pm1)$.
This means that the resonant and pseudo-resonant frequencies obtained
from (\ref{average_M}) {[}or (\ref{average_final}){]} must
be divided by 2 to obtain the correct 
resonant frequencies in
a ZL. Respectively, now one considers the Fourier expansion of $\tmu_{\alpha}(k)$
over the extended BZ. Since profiles of the bands of
ZLs in the extended BZ are more complex than those of OLs,
one usually has to use three harmonics in the Fourier expansion for
their accurate description. Keeping three harmonics in this expansion
($M=3$), we computed the corresponding resonance frequencies for
ZLs that are summarized in Table~\ref{tab:two}. Pseudo-resonant frequencies
are now given by $\tilde{\omega}/2$. It should be stressed here that
even though the first and second bands cross each other {[}Fig.~\ref{fig:one}(c){]},
the coefficients in their expansion as a Fourier series for extended
BZ are different and, therefore, the resonant frequencies
obtained from (\ref{average_M}) for the first and second bands are
different too. This is readily apparent from Fig.~\ref{fig:six} that
illustrates completely different shapes and structure of resonances
for the first and second band excitations.

The Table~\ref{tab:two} has a new column compared with Table~\ref{tab:one},
which is the induced effective dispersion calculated using the estimate
(\ref{estim_width}). Notice that in contrast to ZLs, in OLs all resonances
correspond to strongly suppressed effective dispersion and to good
localization of the wavepacket, and therefore effective dispersion
is not included into Table I. Because of the relatively strong assumptions
made upon calculation of the effective dispersion, the figures presented
in Table II should be considered as approximations; they adequately
characterize the dispersion after only one period of oscillations. Furthermore,
since Eq.~(\ref{estim_width}) is not applicable to pseudo-resonances,
the respective cells are left empty.

\begin{figure*}[t]
\includegraphics[width=1\textwidth]{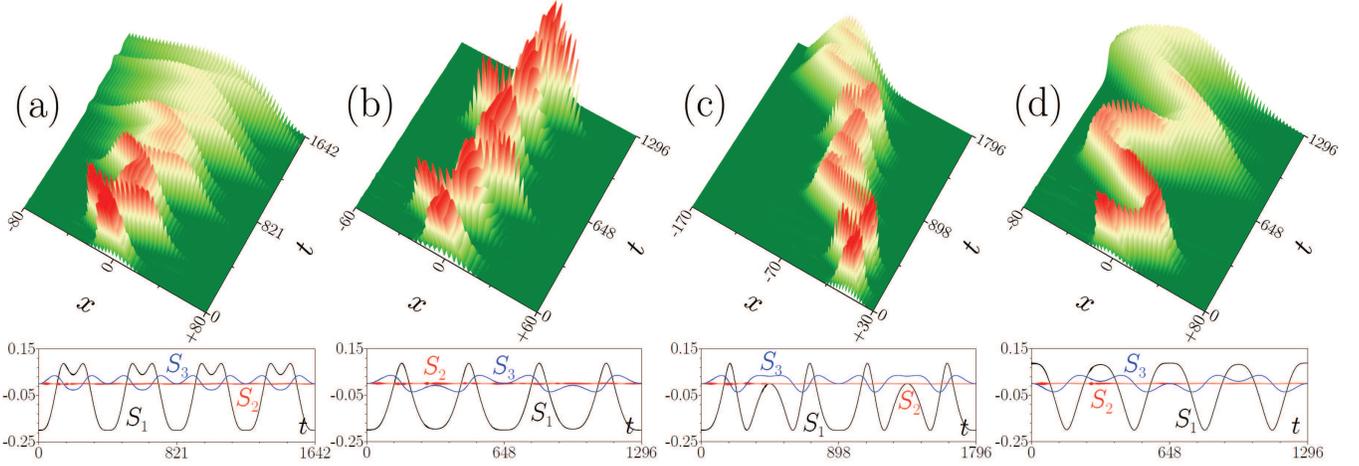} 
\caption{(Color online) Evolution dynamics for the first-band excitations at
(a) $\omega=0.0153$, (b) $\omega=0.0097$, (c) $\omega=0.0070$,
and of the second-band excitation at (d) $\omega=0.0097$ in the ZL
at $\gamma=2$. Only $|\psi_{1}|$ component is shown. Second row
shows corresponding expectation values of the spin components. In
all cases $d(t=0)=7\pi$. In panels (b)-(d) evolution time corresponds
to two complete oscillation periods, while in (a) it corresponds to
eight oscillation periods. }
\label{fig:seven}
\end{figure*}

We observe that the strength of the induced effective dispersion in
ZLs may differ by an order of magnitude for different resonances. This
allows us to classify resonances with suppressed dispersion $\delta d(T)\sim1$
as {\em high-quality} ones, while resonances with $\delta d(T)\gtrsim10$
are {\em low-quality} resonances, where after only a couple of periods one
observes delocalization of the wavepacket even though wavepacket
displacement does not occur. In particular, for $\gamma=0$ high-quality resonances
are found only for the first-band excitations, while all resonances
for the second-band excitations are associated with strong dispersion.
In the case $\gamma=2$ the situation is somehow opposite: two
highest resonances for the second-band excitation correspond to good
localization, while for the first-band excitations only two out of
six highest resonances are high-quality. Importantly, all predicted resonant
frequencies as well as the orders of magnitude of the effective dispersion
(directly manifested in the final width of the wavepacket) are confirmed
numerically, as shown in  Figs.~\ref{fig:five}--\ref{fig:seven}.

In order to show that the dispersion of the resonances
strongly varies with an increasing SO-coupling strength, in
Fig.~\ref{fig:five}(c) we plot the output width of the wavepacket
calculated in two outermost right resonances from Fig.~\ref{fig:six}(a)
(marked with red and black points). The wavepacket width in the authentic
resonance (curve with red dots) gradually grows with $\gamma$, but
decreases in pseudo-resonance (curve with black dots),
so that at sufficiently large values of $\gamma$ pseudo-resonance
appears to be ``better'' than the authentic one. Note also a significant difference in the resonant curves for the first- and
second-band excitations in Fig. \ref{fig:six}(a), which is explained
by the different values of the Fourier harmonics in the series describing the
shapes of the respective bands. Therefore, the induced
effective dispersion can be potentially observed not only with two-level
atoms in ZLs, but also in one-component systems, provided that they
have band shapes that require for their accurate description more
than one Fourier harmonic.

\begin{table}
\begin{tabular}{c | c|c|c|}
\hline
\multicolumn{1}{|c|}{\multirow{1}{*}{$\gamma$} }  & band & Fourier coefficients & $\omega_{3harm}$ \quad $\omega_{num}$  \quad $\delta d(T)$
	\\
  \cline{1-4}
	\multicolumn{1}{ |c  }{\multirow{2}{*}{\begin{tabular}{l}
				 	\\ 	\\ 0 	\\ 	
			\end{tabular}} } &
	\multicolumn{1}{ |c| }{1} & \begin{tabular}{l}
		$\tmu_{1}^{(1)}\approx -0.2831$ 	\\ 	$\tmu_{1}^{(2)}\approx-0.0464$ 	\\ 	$\mu_{1}^{(3)}\approx 0.0074$
	\end{tabular} &
	\begin{tabular}{ccc}
		0.01169 & 0.0117 & 0.90000
		\\
			\cellcolor[gray]{0.9} 0.00936 & 0.00937 & --
		\\
		0.00522 & 0.00521 & 0.71442
		\\
	\cellcolor[gray]{0.9} 0.00468 &  0.00469 & --
	\end{tabular}
	\\ \cline{2-4}
	\multicolumn{1}{ |c  }{}                        &
	\multicolumn{1}{ |c| }{2} &\begin{tabular}{l}
		$\tmu_{2}^{(1)}\approx 0.2831$
		\\
		$\tmu_{2}^{(2)}\approx -0.0464$
		\\
		$\tmu_{2}^{(3)}\approx -0.0074$
	\end{tabular}
	&
	\begin{tabular}{ccc}
		0.01364 & 0.01359 & 20.74133
		\\
		\cellcolor[gray]{0.9} 0.00936 & 0.00937 & --
		\\
		0.00557 & 0.00557 & 20.70126
		\\
      \cellcolor[gray]{0.9} 0.00468 & 0.00469 & --
	\end{tabular}
	\\ \cline{1-4}
		\multicolumn{1}{ |c  }{\multirow{2}{*}{\begin{tabular}{l}
					\\ 	\\  2	\\ 	
				\end{tabular}} } &
				\multicolumn{1}{ |c| }{1} & \begin{tabular}{l}
					$\tmu_{1}^{(1)}\approx -0.14203$ \\ 	
					$\tmu_{1}^{(2)}\approx 0.06511 $ \\ 	
					$\tmu_{1}^{(3)}\approx -0.00515$
				\end{tabular} &
				\begin{tabular}{ccc}
                0.04689  &  0.04822 & 21.09505
					\\
			     0.01524  & 0.01525 & 28.42983
					\\
					0.00969 &  0.00970 & 2.85704
					\\
				\cellcolor[gray]{0.9} 0.00936 &  0.00936 & --
					\\
                0.00794 &  0.00797 & 23.55634
					\\
				0.00581	 & 0.00583 & 28.48675
					\\
				 0.00476 &   0.00473 & 2.86665
 				\end{tabular}
				\\ \cline{2-4}
				\multicolumn{1}{ |c  }{}        &
				\multicolumn{1}{ |c| }{2} &\begin{tabular}{l}
					$\tmu_{2}^{(1)}\approx 0.14204$
					\\
					$\tmu_{2}^{(2)}\approx 0.06511$
					\\
					$\tmu_{2}^{(3)}\approx 0.00515$
				\end{tabular}
				&
				\begin{tabular}{ccc}
				0.01099 & 0.01095 & 3.48682
					\\
					\cellcolor[gray]{0.9} 0.00936 & 0.00936 & --
					\\
					0.00506 & 0.00505 & 3.55544
					\\
                   \cellcolor[gray]{0.9} 0.00468 & 0.00468 & --
				\end{tabular}
\\ \cline{1-4}
\end{tabular}
	\caption{
		Coefficients of the Fourier expansion of the energy over the extended BZ for ZLs; resonance and pseudo-resonance (in shadowed cells) frequencies $\omega_{3harm}$ computed using three Fourier harmonics, the frequency $\omega_{num}$ obtained from the direct numerical simulations (see Fig.~\ref{fig:six}), as well as the estimate of the induced effective dispersion according to formula (\ref{estim_width}). Since the last formula generically is not valid for pseudo-resonances, the respective cells are left empty.}
	\label{tab:two}
\end{table}

Finally, in Fig.~\ref{fig:seven} we show examples of the dynamics
of Bloch wavepackets in ZLs subject to the periodic linear force. One
observes more complex dynamics inside one period of oscillations, as compared to the case of OLs depicted in Fig. \ref{fig:four}, which is related to the
adiabatic motion of the wavepacket along the first and then along
the second bands. The spin dynamics is different too. In the case of
OLs the average spin exhibits a nearly constant and dominating $x$-component,
in ZLs it rotates in the $(x,z)$ plane {[}the component $S_{2}$ is
nearly negligible, as described in Sec.~\ref{sec:OL} for OLs and that remains valid for ZLs{]}.

\section{Conclusions}

\label{sec:concl}

We addressed the phenomenon of dynamic localization
of a spin-orbit-coupled two-level atom described by a spinor wavefunction,
in optical and Zeeman lattices subjected to a periodically oscillating
linear force. The analytical description of the phenomenon revealed a number of new features,
some of which appear only for spinors while others were never
discussed even in scalar one-component systems. The latter include the dependence of the dynamics
on the phase of the external force, which may give rise to pseudo-resonances.
Such pseudo-resonance frequencies yield zero expectation values for
the atom coordinate over one period. They disappear for specific phases
of the oscillating force, do not depend on the strength of the spin-orbit
coupling, and usually are associated with strong dispersion of the
wavepacket.

We also found that for an arbitrary initial phase
of the periodic force, a deviation of the band structure from the tight-binding
limit (single harmonic in the Fourier spectrum) leads
to induced effective dispersion. A practical consequence of this effect
is the absence of complete localization, i.e.,   restoration of the
initial shape of the wavepacket after one complete oscillation period
of the linear force. It is worth mentioning that this phenomenon is not captured by the tight-binding
approximation.

The effective dispersion allows distinguishing between high-quality and low-quality
resonances: in the former, if the dispersion is strongly suppressed then one observes
almost complete restoration of the initial shape of the wavepackets
after several periods of oscillations, while in the latter resonances
correspond to strong dispersion, even
though the center of spinor wavepacket does not experience displacement.
Since this phenomenon is determined by the deviation of the energy,
i.e., $\mu_{\alpha}(k)$ dependence, from a cosine profile, it is
strongly affected by the spin-orbit coupling that induces strong deformations
of the band-gap structures.

The dynamics of the center of the wavepacket in the optical lattices
does not depart significantly from teh behavior known for one-component wavefunctions,
although now it is accompanied by rotation of the pseudo-spin.
In the case of Zeeman lattices, the conclusion is drastically different.
Indeed, bands in the Zeeman lattices cross each other and the spin of an wavefunction
flips after one adiabatic passage of the Brillouin zone. Only after
the second passage of the Brillouin zone, the spinor returns to its initial state.
Therefore now the resonant frequencies are determined by the time
necessary for the wavefunction to pass two Brillouin zones, rather than one.
In all considered cases the numerically simulated dynamics agrees remarkably well with
the analytical results.

\acknowledgments

The work of VVK and DAZ was supported by the FCT (Portugal) grant
UID/FIS/00618/2013. YVK and LT have been partially supported by the
Severo Ochoa Excellence program and by Fundacio Cellex and Fundacio Mir-Puig Barcelona.

\end{document}